\documentclass[aps,preprint,endfloats*,showpacs,amsfonts,amsmath]{revtex4}

\usepackage{graphicx}  

\usepackage{color}
\begin{document}
\author{M.\  J.\ Greenall}
\affiliation{Institute of Mathematics, Physics and Computer Science,
  Physical Sciences Building, Aberystwyth University, Aberystwyth SY23
  3BZ, United Kingdom.}
\altaffiliation{School of Mathematics and Physics, University of
  Lincoln, Brayford Pool, Lincoln LN6 7TS, United Kingdom.}
\email{mjgreenall@physics.org}
\title{Disk-shaped bicelles in block copolymer/homopolymer blends}
\begin{abstract}
Mixtures of micelle-forming and lamella-forming amphiphiles in solution
can form disk-shaped bilayers, sometimes referred to as bicelles. Using
self-consistent field theory (SCFT), we investigate the structure and
stability of these aggregates
in a blend of two species of PS-PDMS diblock with PDMS
homopolymer at $225^\circ\text{C}$. We find that the center of each
disk is mainly composed of lamella-forming diblocks, while its thicker
rim is mostly formed of micelle-forming diblocks. However, this
segregation is not perfect, and the concentration of micelle formers
is of the order of $10\%$ on the flat central surface of the
bicelle. We also find that the addition of micelle
former to the mixture of lamella former and homopolymer is necessary
for disk-like bicelles to be stable. Specifically, the free
energy density of the disk has a minimum as a function of the disk
radius when both micelle- and lamella-forming diblocks are present,
indicating that the bicelles have a preferred, finite radius. However,
it decays monotonically when only lamella former is present,
indicating that the bicelle structure is always unstable with respect
to further aggregation in these systems. Finally, we identify a concentration range
where the bicelle is predicted to have a lower free energy density
than the simple cylindrical and lamellar aggregates, and so might be
thermodynamically stable.
\end{abstract}
\maketitle
\section{Introduction}
Mixtures of lamella-forming and micelle-forming amphiphiles can form structures in solution that are not
seen when only one type of amphiphile is present
\cite{jain_bates_macro,safran_pincus_andelman, zidovska}. These structures, which
include disks, ribbons, and perforated lamellae \cite{duerr,katsaras}, have been
studied in particular depth in lipid-detergent mixtures, where they are known
collectively as \textit{bicelles} \cite{duerr_chem_rev}. Bicelles are
widely used in biophysical experiments
as model membranes \cite{duerr} for the solubilization of 
proteins \cite{sanders_landis,seddon}, and have
several advantages over alternative membrane structures such as multilayers and
micelles. In particular, they avoid the curvature and strain effects
that can occur in micelles \cite{matsumori,prosser}, and, unlike oriented
multilayers, are usually optically
tranparent \cite{sanders_prosser}. They contain smaller amounts of
detergent, which can be harmful to membrane proteins,
than mixed micelles \cite{seddon}, and, under certain circumstances
\cite{loudet,diller,macdonald,duerr,ram_prestegard}, can be aligned in a magnetic
field. Which of the various bicellar morphologies forms depends on a
variety of factors \cite{sanders}, with the disk morphology appearing
at lower temperatures \cite{ramos} and lower lamella-former
concentrations \cite{duerr}, and perforated lamellae forming at higher
temperatures \cite{ramos} and higher concentrations of lamella former \cite{duerr}.

Much progress has also been made in the modeling of
bicelles, using techniques ranging from simple geometric models
\cite{vold_prosser} to microscopic simulations \cite{vacha}. Recently, a theory
based on the chemical potentials of the amphiphiles in differently
shaped aggregates has been used to model light-scattering data on disk
formation in mixed surfactant systems \cite{anachkov}. Disk-shaped bilayers and ribbon-like structures have been
found in coarse-grained molecular dynamics simulations
\cite{dejoannis,noguchi2,lee_pastor,vacha}, and similar techniques have been used to investigate
the transformation of small vesicles into disks
\cite{shinoda,noguchi} and vice versa \cite{marrink_mark}. However,
the complexity of lipid-detergent systems, and the simplifying
assumptions that must be made when modeling them, mean that detailed
comparison of theoretical models with experimental results can prove difficult.

Fortunately, it can sometimes be possible to perform more detailed comparisons of
theory with experiment for the self-assembly of micelles and bilayers
in systems that are composed of a diblock copolymer and a homopolymer,
rather than a lipid or detergent dissolved in water \cite{kinning_thomas_fetters,roe,leibler_orland_wheeler,mayes_delacruz}. In these
systems, scaling theories
\cite{leibler_orland_wheeler,roe} and self-consistent
field theory \cite{gbm_macro} give a good description of the
size of the micelles, and self-consistent field theory can also
predict the shape of the aggregates that will be formed for a given
set of polymer parameters \cite{gbm_jcp}.

In the current paper, we take a similar approach to the study of disk-shaped
bicelles, and use self-consistent field theory \cite{edwards} to investigate the structures formed in a blend of two
types of 
poly(styrene)-poly(dimethylsiloxane) (PS-PDMS) copolymer with
poly(dimethylsiloxane) (PDMS) homopolymer. We have a number of reasons
for focusing on this system. First, the two polymers segregate
strongly \cite{cochran_morse_bates,phys_prop_poly}, with the result
that the system will mimic the lower temperatures where the
disk-shaped bicelle morphology is most likely to be seen. Second, the
large difference in electron absorption and scattering of PS and PDMS
means that the microstructures formed in this system can be studied
directly by electron microscopy, without the need for staining
\cite{saam_fearon}. The third reason is the possibility of modifying
the properties of the PDMS. The two polymers have very different glass
transition temperatures \cite{chow,krause_lu_iskandar}, and the PS blocks
will be become glassy when the system is cooled to room temperature, while
the PDMS will remain viscoelastic, yielding a dispersion of hard disks. Such systems
\cite{dykes} can display a range of behavior, including
shear-induced phase separation \cite{brown_rennie} and the formation
of networks of platelets \cite{nicolai,patil}.

The paper is organized as follows. In the following section, we describe our PDMS/PS system. Next, we give a brief
introduction to the technique to be used, self-consistent field
theory (SCFT). We then present and discuss our results,
and give our conclusions in the final section.

\section{Details of the system}\label{details_of_system}
As discussed in the Introduction, blends of PS and PDMS are
strongly segregated, and the Flory $\chi$ parameter between the two
species is high. It is given as a function of temperature $T$ (measured in
K) by \cite{cochran_morse_bates,phys_prop_poly} $3.1\times
10^{-2}+58/T$. This expression for
$\chi$ is defined with respect to a reference volume of
$1/\rho_0=100\text{\AA}^3$, and is valid for temperatures from 165 to
225$^\circ$C. To keep our SCFT algorithm numerically stable, we carry
out our calculations at the upper end of this range (225$^\circ$C),
where the polymers will be slightly more weakly segregated and the
interfaces less sharp. The lamella-forming species is chosen to have
a PDMS block of molar mass 6000g/mol and a PS block of 13000g/mol, and
the molar mass of the PDMS homopolymer is set to 4500g/mol. The 
homopolymer has been chosen to be shorter than the PDMS block of the
copolymer to ensure thermodynamic stability
\cite{mayes,leibler_orland_wheeler}, and the ratio of the block
lengths of the copolymer has been set so that this species will have a
strong tendency to form lamellae
\cite{kinning_winey_thomas,gbm_jcp}. A polymer chain with
`hydrophobic' and `hydrophilic' components in a ratio of roughly two
to one, as is the case here, also provides a simple model of the long-chain DMPC molecules
that form the body of bicelles in lipid-detergent mixtures
\cite{duerr,vacha}. To ensure that it has a clear preference for
forming micelles, the other species has a long PDMS block,
with molar mass 24000g/mol, and a PS block of 6000g/mol. The overall
weight fraction of copolymer, including both species, is set to
$2.5\%$. Using this relatively low concentration has two advantages. First, it will allow us to neglect interactions
between the aggregates when modeling the thermodynamics of the
system. Second, it will avoid interference between the aggregate and
the boundary of the calculation box. We do not risk dropping below the
critical micelle concentration by using a copolymer concentration of
this magnitude, because the system under
consideration is very strongly segregated\cite{kinning_thomas_fetters}. We also need to know
the mean-square end-to-end distance, $r_0^2$, of each polymer species as a
function of its molar mass, $M$. For polymers in a melt, this is
given by an expression of the form $r_0^2\propto M$, where the ratio
$r_0^2/M$ is close to constant for a given polymer
\cite{polymer_handbook}. For PS \cite{polymer_handbook}, we have that
$r_0^2/M\approx 0.49$. The dimensions of PDMS \cite{konishi} are
rather similar, and $r_0^2/M\approx 0.53$. 

Certain of the parameters above cannot be used directly as input
to an SCFT calculation, and need to be converted into the appropriate
forms. First, the molar masses listed above are converted to molar
volumes \cite{kinning_thomas_fetters} using the specific volumes (in $\text{cm}^3/\text{g}$) of PS
and PDMS. Empirical expressions for the specific volumes of PS
\cite{richardson_savill} and PDMS \cite{shih_flory} as a function of
temperature are taken from the literature. The molar volumes are then used to
calculate the volume fraction of PDMS in each type of diblock and the volume ratios
of the various species. The overall weight fraction of diblocks is
converted to a volume fraction by a similar procedure, and the volume of a
single molecule of each species in $\text{\AA}^3$ is calculated by
dividing the appropriate molar volume by $0.60221413$, a numerical
constant that incorporates Avogadro's number and the conversion from
$\text{cm}^3$ to $\text{\AA}^3$. Finally, the number, $N$, of repeat
units in a polymer chain can be calculated by dividing the volume of
the molecule by the reference volume, $1/\rho_0$.

\section{Self-consistent field theory}\label{scft}
Self-consistent field theory (SCFT) \cite{edwards} is a
mean-field model that has been used
with success to calculate the form and free energy of equilibrium
\cite{maniadis,drolet_fredrickson,matsen_book} and metastable
\cite{duque,katsov1} structures in systems composed of homopolymers \cite{werner}, copolymers
\cite{mueller_gompper,wang} and mixtures of these \cite{denesyuk}. SCFT has
several features that make it especially suited to the current
problem. First, as stated above, it has been shown to give a good description of the shape and size
of micelles in blends of a single species of block copolymer with a
homopolymer \cite{gbm_macro,gbm_jcp}, a system closely related to the
current one. Second, it is faster than simulation
techniques such as Monte Carlo, but can yield comparably accurate
predictions of micelle size and shape
\cite{cavallo,wijmans_linse,leermakers_scheutjens-shape}. This will
allow bicelles of a range of sizes to be studied in a reasonable
period of time. Finally, it
does not require any assumptions to be made regarding the segregation
of
copolymers of different architecture within an aggregate, meaning that
any such effects that we observe are a natural prediction of the theory and have not
been added by hand.

We now give a short overview of the application of SCFT to our system of two
copolymers and a homopolymer, and refer the reader to reviews
\cite{matsen_book,fredrickson_book,schmid_scf_rev} for in-depth
presentations of the theory
and to earlier papers \cite{gg,gbm_jcp,gbm_macro} for a detailed description of our calculations. SCFT is a coarse-grained theory, and individual
molecules are modeled as random walks
in space \cite{schmid_scf_rev}. An ensemble
of many such molecules is considered, and the intermolecular
forces are modeled by introducing contact potentials between
the molecules and assuming that the blend is incompressible \cite{matsen_book}. The
Flory $\chi$ parameter discussed above is used to specify the strength of the repulsion between the two chemical species. In order to reduce
the computational difficulty of the problem, fluctuations are
neglected; that is, a mean-field
approximation is made
\cite{matsen_book}. In the case of
long molecules, this
approximation is quantitatively accurate
\cite{cavallo,fredrickson_book,matsen_book}.

SCFT can be used to perform calculations in different thermodynamical
ensembles \cite{duque,zhou}. In this paper, we perform all calculations in the canonical
ensemble, which corresponds to keeping the amounts of copolymer and homopolymer in the
simulation box fixed. This approach
makes it easier for us to access more complex
aggregates, such as bicelles. Such structures are
more difficult to find in ensembles where
the concentrations of the various species are able to change, and
sometimes need to be stabilized by applying
geometric constraints to the density profile \cite{katsov1}.

Applying the mean-field approximation\cite{matsen_book} to our system,
we find that the SCFT expression for the free energy of a system of
copolymer species 1 and 2 in
homopolymer is given by
\begin{align}
\frac{FN_1}{k_\text{B}T\rho_0V}&=\frac{F_\text{H}N_1}{k_\text{B}T\rho_0V}\nonumber\\
&
-(\chi
  N/V)\int\mathrm{d}\mathbf{r}\,[(\phi_\text{PDMS1}(\mathbf{r})+\phi_\text{PDMS2}(\mathbf{r})
  +\phi_\text{hPDMS}(\mathbf{r})-\overline{\phi}_\text{PDMS1}-\overline{\phi}_\text{PDMS2}-\overline{\phi}_\text{hPDMS})\nonumber\\
&\times(\phi_\text{PS1}(\mathbf{r})+\phi_\text{PS2}(\mathbf{r})-\overline{\phi}_\text{PS1}-\overline{\phi}_\text{PS2})]-(\overline{\phi}_\text{PDMS1}+\overline{\phi}_\text{PS1})\ln (Q_1/V)
\nonumber\\
& 
-[(\overline{\phi}_\text{PDMS2}+\overline{\phi}_\text{PS2})/\alpha]\ln
(Q_2/V) -(\overline{\phi}_\text{hPDMS}/\alpha_\text{h})\ln (Q_\text{h}/V)
\label{FE}
\end{align}
where the $\overline{\phi}_i$ are the mean volume fractions of the
various components, with $i=\text{PDMS1}$ or $\text{PDMS2}$
for the poly(dimethyl siloxane)
components of species $1$ and $2$, $i=\text{PS1}$ or $\text{PS2}$ for the poly(styrene)
components and $i=\text{hPDMS}$ for the homopolymer solvent, and the $\phi_i(\mathbf{r})$
are the local volume fractions. $V$ is the
total volume, $N_1$ is the number of repeat units in species 1, and $F_\text{H}$
is the SCFT free energy of a homogeneous system of the same
composition. The architectures of the individual molecules enter through the
single-chain partition functions $Q_{j=1,2,h}$, which are calculated from the propagators $q$ and $q^\dagger$ \cite{matsen_book}. These
latter quantities satisfy diffusion equations with a field term
that incorporates the polymer interactions. One field is associated
with the PDMS segments, and one with the PS segments. This means
that, to calculate the
copolymer partition functions, $Q_1$ and $Q_2$, the diffusion equation
must be solved with the field appropriate to each of the two blocks
of the copolymer \cite{fredrickson_book,matsen_book}. In addition, the
difference between the
expressions for the mean-square end-to-end distances of PDMS and PS
\cite{konishi,polymer_handbook} must be taken into account when
calculating the prefactor of the $\nabla^2q$ term in each block. The polymer density
profiles are computed from integrals over the propagators
\cite{matsen_book,fredrickson_book}, with the volume fractions of PS and
PDMS in each copolymer species entering via the limits of integration.

We perform all our calculations in cylindrical polar coordinates. Since we mainly focus on
disk-shaped aggregates, we reduce the problem to a two-dimensional one
by assuming that
the system has rotational invariance about the $z$-axis, and carry out
all our calculations in a cylindrical box. We impose reflecting
boundary conditions at all edges of the box, and the center of the
bicelle lies at the origin of the coordinate system. The vertical
height of this box is set to $Z=450\text{\AA}$ (meaning that its effective
height is $900\text{\AA}$), and its radius $R$ is varied
according to a procedure that will be described later. The diffusion
equations are solved using a finite difference method
\cite{num_rec} with a step size of $2.5\text{\AA}$. The
curve parameter $s$ that specifies the distance along the polymer backbone\cite{fredrickson_book} runs from $0$ to $1$, and its step size is set to $1/800$ for the copolymers
and $1/100$ for the homopolymer.

The derivation of the mean-field free energy $F$ also yields a set
of simultaneous equations relating the fields
$w_\text{PDMS}(\mathbf{r})$ and $w_\text{PS}(\mathbf{r})$ to the densities
$\phi_i(\mathbf{r})$. To calculate the density profiles for a given set of
mean volume fractions $\overline{\phi}_i$, we make an initial guess
for the fields that has the approximate form of the structure we wish to
study, and then solve the diffusion
equations to calculate the propagators and density profiles
corresponding to these fields. The new $\phi_i(\mathbf{r})$ are then
substituted into the simultaneous equations to compute updated fields \cite{matsen2004}, which are then used in turn to
compute new $\phi_i(\mathbf{r})$ by solving the diffusion
equation as described above. For the algorithm to remain
stable, the iteration needs to be damped, and, instead of using the updated
values of $w_i$ directly to calculate the $\phi_i$, we use the linear combination
$\lambda w_i^\text{new}+(1-\lambda)w_i^\text{old}$ where
$\lambda\approx 0.04$. This process is repeated until
convergence is achieved. For smaller systems, this {\it simple mixing} is
sufficient. However, in larger calculation boxes, it can stall after
an initial period of convergence. When performing calculations on
these systems, we follow Thompson and co-workers \cite{thompson} in
passing the fields generated by the simple mixing iterations to an
Anderson mixing algorithm
\cite{anderson}. This algorithm has greater flexibility in the
iteration steps it can take, since it stores a
history of previous values of the $w_i(\mathbf{r})$, and calculates the next
estimate for the fields by adding these together in a linear
combination \cite{ng}. The pure Anderson mixing procedure is not
stable in our case, and has to be damped
\cite{schmid_pre,stasiak_matsen}. We find that a history of
50--100 previous values of the $w_i(\mathbf{r})$ and damping parameter of
0.1--0.2 yield good results. 

We now need to address the issue of how to relate the thermodynamics
of a single bicelle to those of a larger system containing many
aggregates. To do this, we adapt a procedure that has been developed
to study simple micelles and bilayers
\cite{gbm_macro,gbm_jcp, gm_prl,shull}. First, we
compute the free-energy
density of a cylindrical box containing a single disk-shaped
aggregate. Since the copolymer concentration is low, the aggregate is
surrounded by a large volume of homopolymer, and the shape of the aggregate is
not influenced by contact with the boundaries of the system. The radius of the simulation box is then varied, keeping the total volume fractions
of both types of copolymer
constant. Changing the size of the box in this way causes the bicelle
to grow in the radial direction. The free-energy density of the system is calculated for
each box radius. As is the case when this method is applied to
micelles, there will be a minimum in
the free energy as a function of the radius\cite{shull} if the bicelle is stable. This solution of the SCFT
equations corresponds to the optimum size of the bicelle. If, on the
other hand, the bicelle
is unstable with respect to further aggregation, the free energy
density will decay monotonically as the box radius is increased,
meaning that the bicelle can always
move to a more energetically favorable state by growing in the radial
direction, eventually forming an extended bilayer.

This approach is designed to
mimic the behavior of a larger system (of fixed
volume and fixed copolymer volume fraction) containing many
bicelles. The reason for this is that this larger system minimizes its total
free energy by changing the number of aggregates and hence the volume
(`box size') occupied by each. Minimizing the free energy density in this way
locates the bicelle that is the most energetically favorable and
therefore the
most likely to be observed in a sample containing many aggregates.

This minimum of the free-energy density with respect to the box size
$V$ corresponds to the absolute free energy
minimum of a solution of aggregates, and a point on the curve $F(V)/Vk_\text{B}T$ corresponds to a {\em monodisperse} solution of aggregates of a
given size. In an earlier publication \cite{gm_prl}, we showed how to use these curves to
take into account simple fluctuations around the free energy minimum
and so calculate the width, $\Delta$, of the size distribution of
aggregates in
a system containing only one type of amphiphile. This was achieved by
relating the curvature of  $F(V)/Vk_\text{B}T$
to the second derivative, $\partial^2f_p/\partial p^2$, of the free energy $f_p$ of an aggregate
containing $p$ molecules, which was then used \cite{puvvada} to calculate $\Delta$
via $1/\Delta^2=(1/k_\text{B}T)\partial ^2f_p/\partial p^2$
. We began by writing down an expression for the free-energy density
$F/Vk_\text{B}T$ of a monodisperse system of
aggregates, each containing $p$ copolymers. This was given by
\begin{equation}
\frac{F}{Vk_\text{B}T}=(c-pc_p)\{\ln[(c-pc_P)v_1/e]+f_1\}+c_pf_p
\label{FE_single_species}
\end{equation}
where $c$ is the number density of copolymers, $c_p=1/V$ is the number
density of aggregates, $f_1$ is the free energy of a copolymer in
solution, $f_p$ is the free energy of an aggregate of $p$ copolymers,
$v_1=N_1/\rho_0$ is the volume of a single copolymer, 
and first term arises
from the entropy of the free copolymers in solution.
We then noted that a single SCFT
calculation finds the local minimum of the free energy density $F/Vk_\text{B}T$ for an aggregate in a box
of volume
$V$. In the process, it determines the optimum number of
molecules in the aggregate for this box size and so corresponds to
minimizing $F/Vk_\text{B}T$ with respect to $p$ at a given aggregate
number density $1/V$. Varying
$V$ then yields the curve $F(V)/Vk_\text{B}T$, from which we can
read off $\mathrm{d}^2[F/Vk_\text{B}T]/\mathrm{d}V^2$. Remembering that this derivative
is evaluated along the line where $\partial [F/Vk_{\text{B}}T]/\partial p|_V=0$,
we found that
\begin{equation}
\frac{1}{\Delta^2}=\frac{1}{k_{\text{B}}T}\frac{\partial^2 f_p}{\partial
  p^2}=\frac{v_1}{v_{\text{a}}^2/(V^3\mathrm{d}^2\tilde{F}/\mathrm{d}
  V^2)-(\phi V-v_{\text{a}})},
\label{polydisp}
\end{equation}
where we have written $\tilde{F}=Fv_1/Vk_\text{B}T$,
$v_{\text{a}}=pv_1$, and converted the number density $c$ to the
volume fraction $\phi$ to express $\Delta$ in terms of
quantities that are either input to or results of our SCFT
calculations. 

We now extend this
calculation to a system containing two amphiphile species. As in the
calculation above, we suppose that there is only one type
of aggregate in the system. We make the further simplifying assumption
that the number fractions of the two species in the aggregate are the
same as the overall number fractions. This turns out to hold accurately in our numerical results, and allows us to write the
free-energy density of the two-species system as
\begin{eqnarray}
\frac{F}{Vk_\text{B}T}&=&(\psi_1c-\psi_1pc_p)\{\ln[(\psi_1c-\psi_1pc_P)v_1/e]+f_{11}\}\nonumber\\
&&+(\psi_2c-\psi_2pc_p)\{\ln[(\psi_2c-\psi_2pc_P)v_2/e]+f_{12}\}
+c_pf_p,
\end{eqnarray}
where the number densities of species 1 and 2 are $\psi_1c$ and
$\psi_2c$ respectively (so that $\psi_1+\psi_2=1$), $f_{11}$ and
$f_{12}$ are the free energies of single copolymers of species 1 and 2
in solution, and $v_2$ is the volume of a single copolymer of species
2. This assumption also allows us to write the number of copolymers in
the aggregate as $p=v_\text{a}/v_\text{eff}$, where
\begin{equation}
v_\text{eff}=\frac{\phi_1v_1}{\phi_1+\phi_2v_1/v_2}+\frac{\phi_2v_2}{\phi_1v_2/v_1+\phi_2}.
\label{v_eff}
\end{equation}
The calculation then proceeds as before \cite{gm_prl},
and we find that
\begin{equation}
\frac{1}{\Delta^2}=\frac{1}{k_{\text{B}}T}\frac{\partial^2
  f_p}{\partial p^2}
=\frac{1}{v_1(v_\text{a}/v_\text{eff})^2/(V^3\mathrm{d}^2\tilde{F}/\mathrm{d}V^2)-[(\phi_1/v_1+\phi_2/v_2)V-v_\text{a}/v_\text{eff}]},
\label{polydisp2}
\end{equation}
where, as before, we have converted number densities to volume
fractions. The extra factor of $v_1$ in equation \ref{polydisp2} arises from the
normalization of the free energy in equation \ref{FE}.

To see if the bicelle is likely to form, we also need to compare its
free energy density with those of other aggregates. If we can
establish that, for a given concentration range, the bicelle has a
lower free-energy density than the simple spherical, cylindrical and
lamellar aggregates, there is a clear possibility that it will form in
this region. To calculate the free energy density of 
spherical micelles, we simply continue our calculations to smaller
values of the calculation box radius until the disk-like bicelle
shrinks to a sphere. It turns out that, for the systems considered
here, where the $\chi$ parameter is large and a high proportion of thepolymers have a
strong preference for forming lamellae, spherical micelles have much
higher free-energy densities than the other aggregates. In
consequence, we omit them from our results. The free energies of the optimum cylinder and
lamella are found by a similar method to that described above \cite{gbm_jcp}. These structures
are assumed to be of infinite extent, so calculating their optimum free
energies using SCFT is a one-dimensional problem, with the box size being varied in the $r$- and $z$-directions
respectively\cite{gbm_jcp}. For convenience, and for consistency with our earlier
calculations, we perform these calculations using the same
two-dimensional algorithm as before, but with the calculation box made
very thin in the direction in which the density profiles do not vary.

\section{Results and discussion}\label{results}

We begin this section by demonstrating that the disk-like bicelle structure is a
solution to the SCFT equations. Next, we investigate the dependence of the free energy density
of the disk-like bicelle on its radius, to determine in which systems
the bicelle has a preferred size. Finally, we compare the free energy
of the bicelle to those of the simple cylindrical and lamellar
structures for a range of concentrations of the two copolymer species,
to find the range of parameters for which it might form in an experiment.

\subsection{The disk-shaped bicelle morphology}\label{disks}
\begin{figure}
\includegraphics[width=0.75\linewidth]{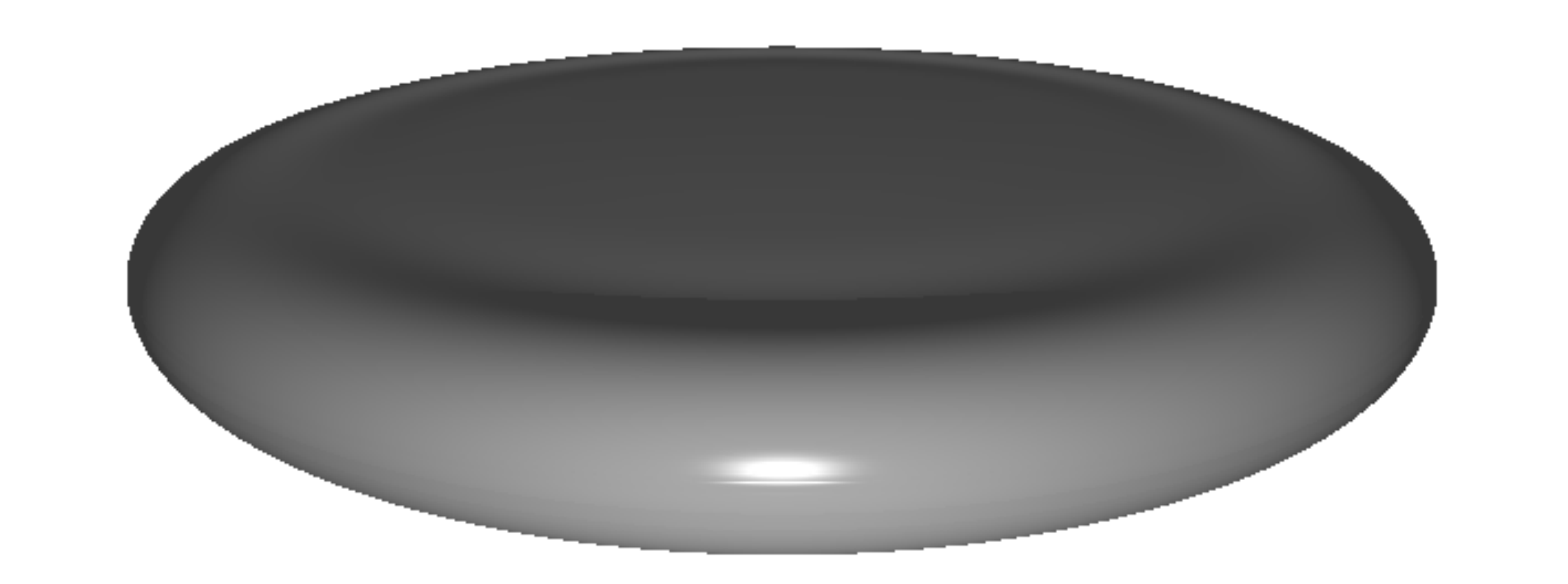}
\caption{\label{lathe_fit_fig} Ray-traced plot of the surface of the bicelle. 
}
\end{figure}

In Figure \ref{lathe_fit_fig}, we show a ray-traced plot of the surface of a disk-like bicelle
obtained as a solution to the SCFT equations in the PS/PDMS system
described above. $30\%$ by weight of diblocks are sphere
formers, and $70\%$ are lamella formers. The surface is defined as the
locus of the points where the local solvent volume fraction
$\phi_\text{S}(\mathbf{r})=0.5$, and has a biconcave disk shape
reminiscent of a red blood cell. 

\begin{figure}
\includegraphics[width=0.75\linewidth]{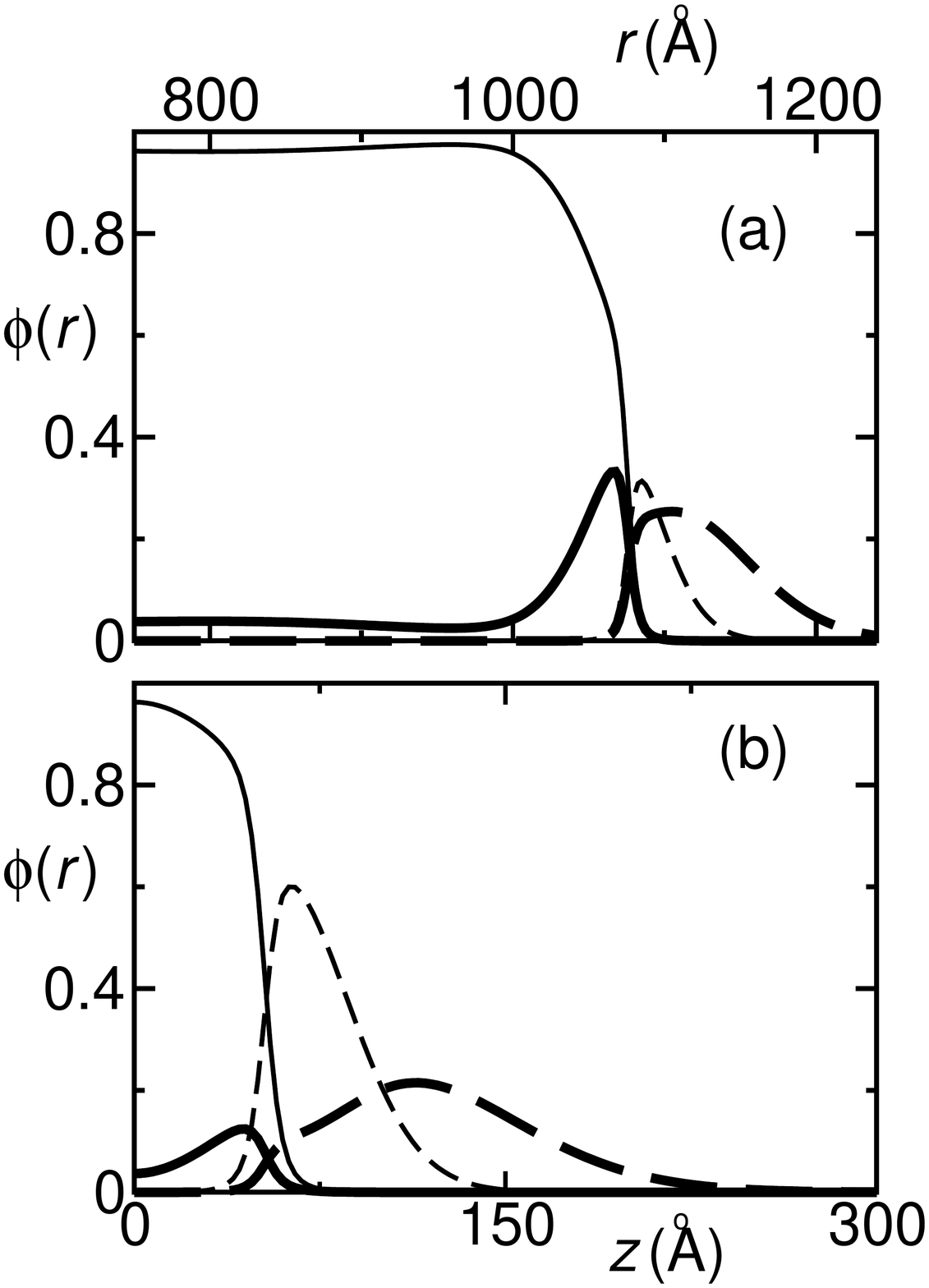}
\caption{\label{outline_profiles_fig} Cuts through the volume fraction
  profiles of the various blocks making up the bicelle: sphere
  former PS blocks (thick solid line); sphere former PDMS blocks
  (thick dashed line); lamella former PS blocks (thin solid line);
  lamella former PDMS blocks (thin dashed line). The volume fraction
  profile of the PDMS homopolymer ``solvent'' is omitted for
  clarity. Panel (a) shows cuts in the $r$-direction, and panel (b)
  shows cuts in the $z$-direction. Note that neither panel shows the
  full range of the calculation box.
}
\end{figure}

We now investigate the distribution of the two species, and their
hydrophilic and hydrophobic components, within the bicelle. In Figure \ref{outline_profiles_fig}a, we show radial cuts through the volume fraction profiles of the
various blocks that make up the bicelle. The division of the bicelle
into a well defined ``hydrophobic'' PS core and ``hydrophilic'' PDMS corona is clearly visible, with the
interface between the two regions being located at around $r=1080\text{\AA}$. 
A marked difference in behavior between the sphere-forming and
lamella-forming species is also seen. Specifically, the
sphere formers are concentrated at the edge of the bicelle, with
the volume fraction profile of their hydrophobic components being sharply peaked
just before the core-corona interface, and their relatively long
hydrophilic components stretching out into the solvent. This
structure, in which the species with a preference for
forming curved membranes segregates to the rim of the bicelle, is often sketched in the literature
\cite{sanders_prosser}, and our results show that it can be reproduced
in explicit calculations.

The curves in Figure \ref{outline_profiles_fig}b are cuts through the same
volume fraction
profiles in the $z$-direction. Again, clear core and corona regions
can be seen. The difference between the two plots lies in the fact
that the peak in the
hydrophobic profile of the sphere-forming species just before the
core-corona interface is markedly less pronounced in the $z$-direction, confirming the clear
preference of the sphere formers for the bicelle rim. However, the
presence of this peak, which attains a maximum height of 0.12, shows
that the segregation of the sphere formers to the rim is not perfect,
with an appreciable concentration of this species remaining near $r=0$,
particularly on the flat surface of the bicelle. Again, this result
agrees with the pictorial model of bicelles often suggested in the
biophysics literature \cite{sanders_prosser}.

\subsection{The preferred radius of the bicelle}

Having shown that the bicelle exists as a solution to SCFT and
investigated the distribution of the two copolymer species within it,
we now study the dependence of its free energy density on its radius,
with the aim of finding whether it has a preferred size. The size of
the bicelle is varied by changing the calculation box radius as
described above, and we plot its free energy density against the radius
of its core. We focus on the core radius
because this is the measure of micelle size that is most easily
measured in experiment \cite{kinning_winey_thomas}. Since the
interface between the core and the corona is sharp, all reasonable
definitions of the core radius will yield similar values, and we
define it here as the value of  $r$ at which the core and corona densities are
equal, so that
$\phi_\text{PS1}(\mathbf{r})+\phi_\text{PS2}(\mathbf{r})=\phi_\text{PDMS1}(\mathbf{r})+\phi_\text{PDMS2}(\mathbf{r})$. 

\begin{figure}[H]
\includegraphics[width=0.75\linewidth]{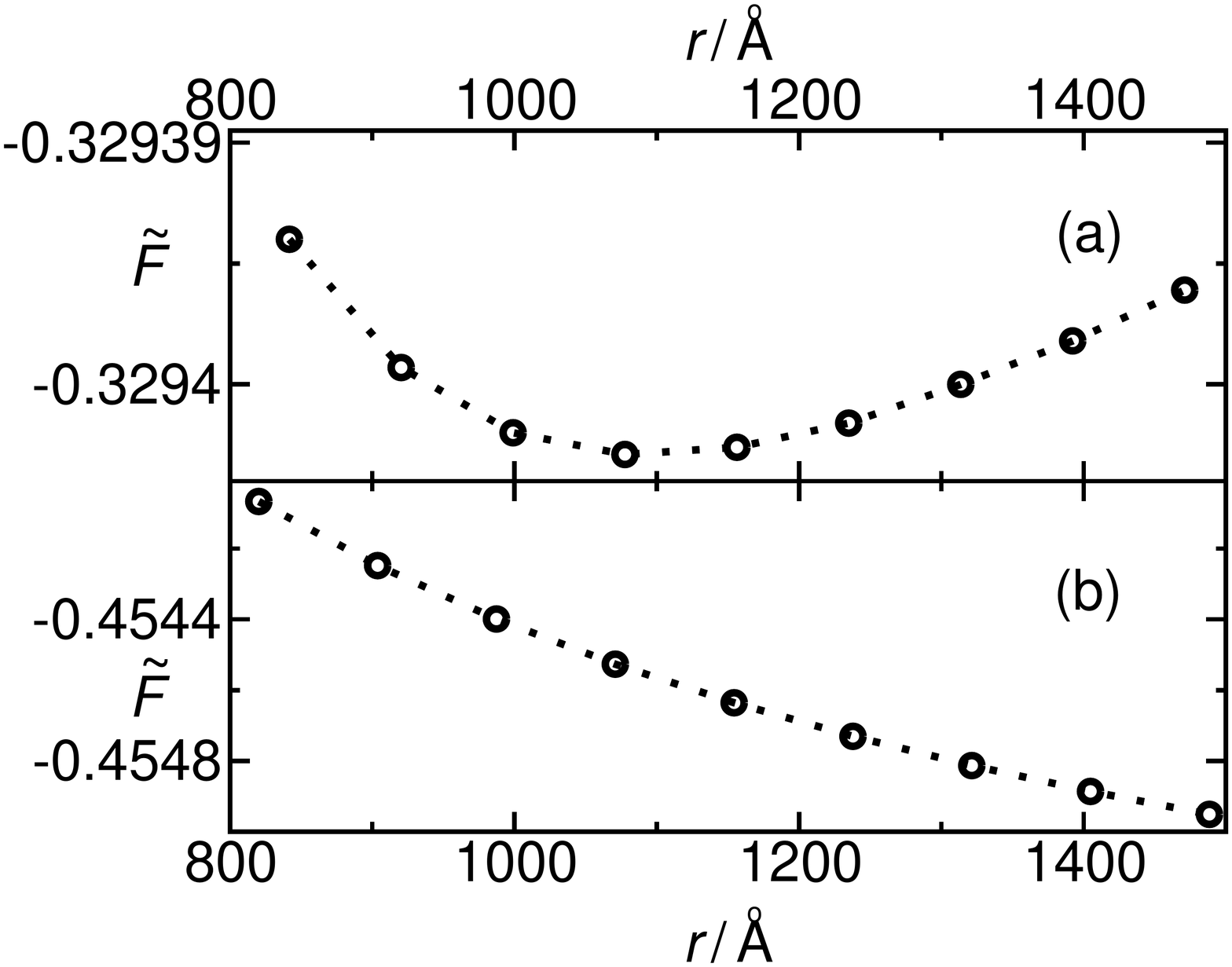}
\caption{\label{mix_vs_sing_fig} Plots of the free-energy density against bicelle core radius for (a) a mixed system in which $30\%$ by weight of all amphiphiles are
sphere formers and $70\%$ are lamella formers and (b) a pure system
in which all amphiphiles are lamella formers. The narrow range of the
$\tilde{F}$ axes arises because the system is dilute and the free energy density is calculated
for the entire calculation box, with the result that the differences in free energy
density between one size of bicelle and another appear small. However,
a more detailed calculation, in which the curvature of the free energy per
chain in the bicelle is extracted from the above curves, reveals that the size selection in this system is
significant, with a relative polydispersity, $\Delta/p$, of approximately $20\%$ in
the aggregation number, corresponding to a relative polydispersity
of around $10\%$ in the radius.
}
\end{figure}

Figure \ref{mix_vs_sing_fig}a shows the results of this calculation for the
system studied above, where $30\%$ by weight of all amphiphiles are
sphere formers and $70\%$ are lamella formers. The curve of
the free energy density shows a clear minimum as a function of the
bicelle radius, showing that the aggregate has a preferred size.

We can now use Equation \ref{polydisp2} to estimate the relative
polydispersity in the aggregation number of the micelle,
$\Delta/p$. This is found to be approximately $20\%$, corresponding
\cite{gm_prl} to a relative polydispersity of around $10\%$ in the
bicelle radius. This demonstrates that significant size selection
takes place in this system, with a clear preferred size for the
bicelles.

\subsection{Concentration dependence}

\begin{figure}[H]
\includegraphics[width=0.75\linewidth]{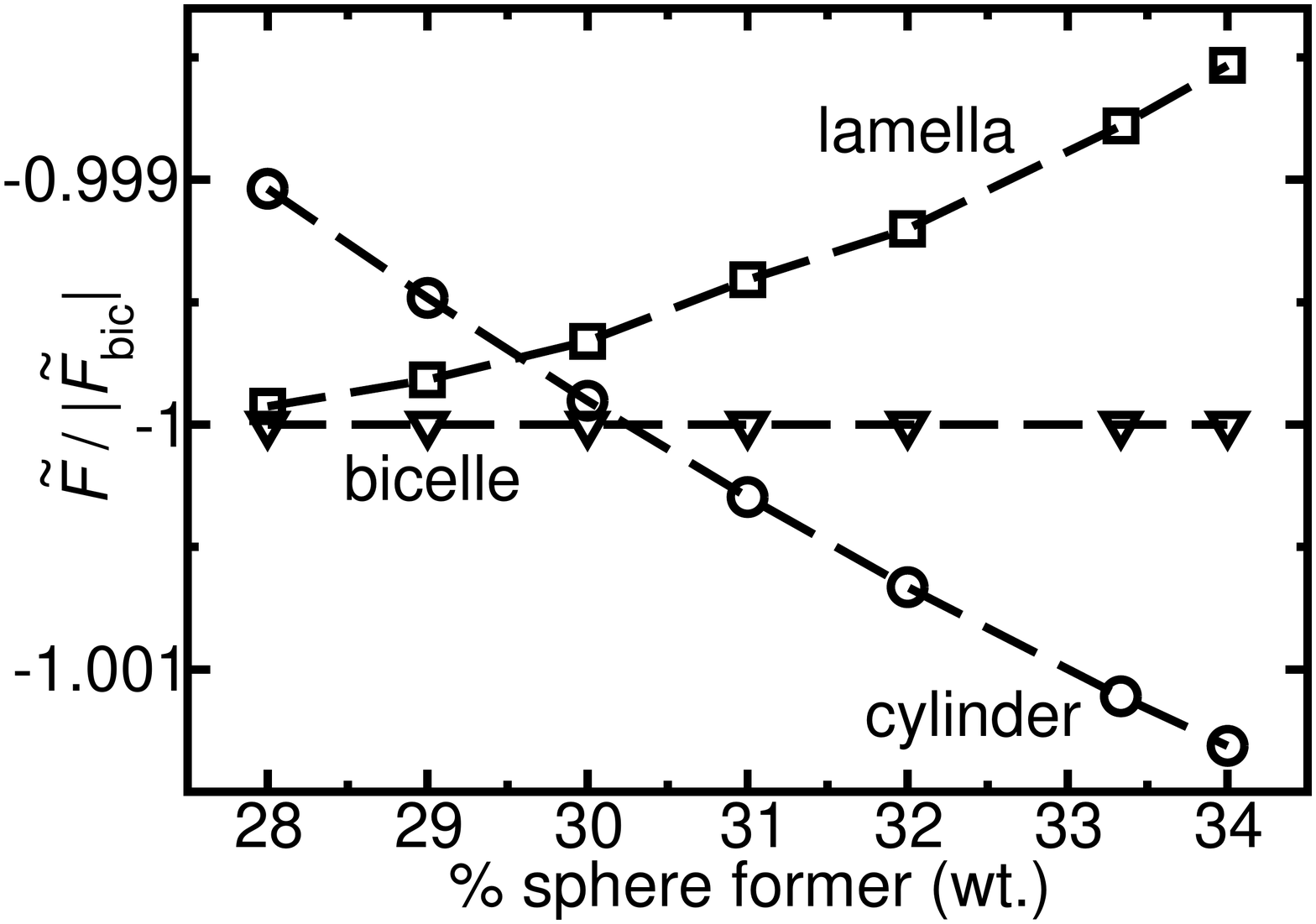}
\caption{\label{FE_norm_fig} Free energy densities of the lamella
  (squares), bicelle (triangles) and cylinder (circles), normalised
  with respect to that of the bicelle and plotted against the percentage
  of amphiphiles that are sphere formers.
}
\end{figure}

\begin{figure}[H]
\includegraphics[width=0.75\linewidth]{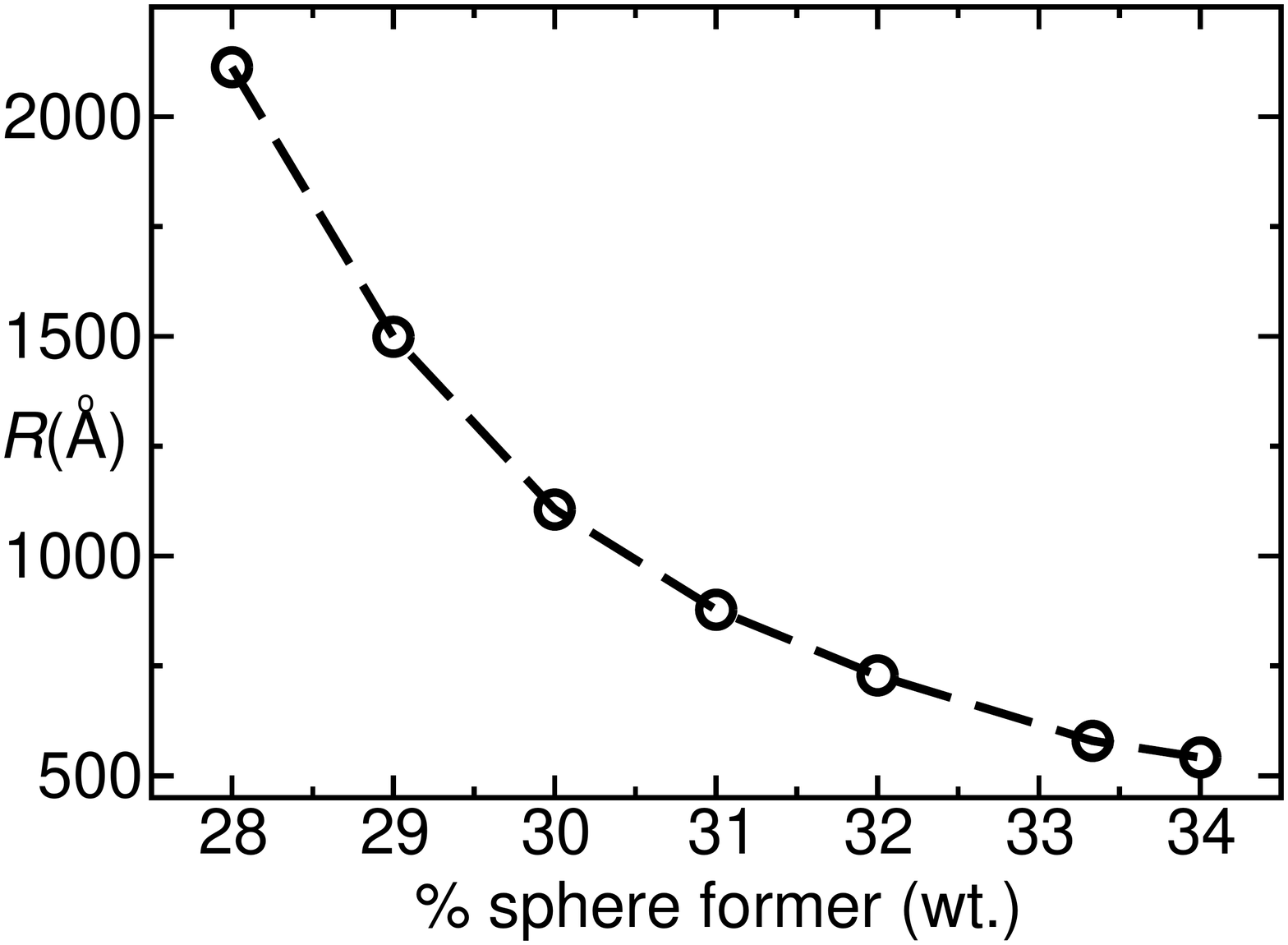}
\caption{\label{radius_fig} The bicelle core radius plotted against
  the percentage of amphiphiles that are sphere formers.
}
\end{figure}

We now compare the free energy density of the bicelle with those of
the cylindrical and lamellar morphologies over a range of
concentrations. Since the bicelle is a hybrid structure that contains elements
of both the lamella and the cylinder, we expect \cite{vacha} that it
will form at compositions around that at which the free energy
densities of these two structures are the
same. We therefore begin by locating this composition, and focus
our attention on its vicinity. Figure \ref{FE_norm_fig} shows the
results of these calculations. At higher weight fractions of
sphere-forming amphiphile, the cylindrical micelle has the lowest
free-energy density and the lamella has the highest. As the
amount of sphere former is reduced, the free-energy density of the
bicelle becomes the lowest of the three aggregates: it drops below that of the cylinder, while remaining below that of
the lamella.
As the weight fraction of sphere former is decreased
further, the free-energy densities of the bicelle and lamella become
closer, although they do not cross within the range of compositions
that we are able to study. It is possible that the two quantities
approach each other asymptotically as the amount of sphere former
tends to zero. This is consistent with the growth in the bicelle
radius seen as the amount of sphere former is reduced (Figure \ref{radius_fig}). As the bicelle becomes very large, it approaches
the lamella in shape, and the free-energy densities of the two
structures can also be expected to become very close.

\section{Conclusions}\label{conclusions}
Using
self-consistent field theory, we have studied the structure and
stability of disk-like bicelles
in a blend of lamella-forming and sphere-forming PS-PDMS diblocks with PDMS
homopolymer. We have found that these structures have a characteristic
biconcave disk shape, like red blood cells. Furthermore, we have shown
that the two amphiphile species are unequally distributed within the
aggregate, with the center being
mainly composed of lamella formers, and the
rim being mostly formed of micelle formers. This picture of bicelles
is often sketched in the biophysics literature \cite{sanders_prosser},
and is reproduced here in calculations on a well defined model.

We also find that the presence of micelle former is necessary
for disk-like bicelles to be stable, and that they will be unstable
with respect to further aggregation when only lamella former is
present. Finally, we locate a region of parameter space range
where the bicelle is predicted to have a lower free energy density
than the competing cylindrical and lamellar aggregates.

There are are a number of ways in which our calculations could be
extended. First, we used the rather high temperature of
$225^\circ\text{C}$ in order to keep the $\chi$ parameter at a level
that our existing numerical methods were able to deal with. Further
refinements of the Anderson mixing method and the algorithm used to
solve the diffusion equation might allow the calculations to be
extended to lower temperatures, where the concentration range at which
the bicelle has a lower free-energy density than the other structures
should be broader. Second, perhaps using these extended methods, we
could look for other polymer blends in which the bicelle is predicted to
be stable and in which its presence could have an effect on the
mechanical properties. Finally, a similar SCFT approach could be applied to lipid
systems \cite{kik}, where bicelles were originally observed. This
could be used to investigate the degree of detergent penetration into
the center of the bicelle, an important issue for the biophysical
experiments that use bicelles as model membranes.

\end{document}